\documentclass[10pt,aps,prl,preprintnumbers,superscriptaddress,showpacs,
twocolumn,floatfix,nofootinbib]{revtex4-2}
\usepackage{graphicx}
\usepackage{amsmath}
\usepackage{slashed}

\newcommand{\gsim}{\;\rlap{\lower 3.5 pt \hbox{$\mathchar \sim$}} \raise 1pt
\hbox {$>$}\;}
\newcommand{\lsim}{\;\rlap{\lower 3.5 pt \hbox{$\mathchar \sim$}} \raise 1pt
\hbox {$<$}\;}

\begin{document}

\title{\boldmath
Majorana Modes of Giant Vortices
\unboldmath}
\author{Logan Gates}
\email[]{lgates@ualberta.ca}
\affiliation{Department of Physics, University of Alberta, Edmonton, Alberta T6G
2J1, Canada}
\author{Alexander A. Penin}
\email[]{penin@ualberta.ca}
\affiliation{Department of Physics, University of Alberta, Edmonton, Alberta T6G
2J1, Canada}
\begin{abstract}
We study Majorana zero modes bound to giant vortices  in
topological superconductors or topological insulator/normal
superconductor heterostructures. By expanding in
inverse powers of a large winding number $n$, we find an
analytic solution for asymptotically all $n$ zero modes
required by the index theorem. Contrary to the existing
estimates, the solution is not pinned to the vortex  boundary
and is composed of the warped lowest Landau level states. While
the dynamics which shapes the zero modes  is a subtle
interference of  the  magnetic effects and Andreev reflection,
the solution is very robust  and is determined by a single
parameter, the vortex radius. The resulting local density of
states has a number of features  which give remarkable
signatures for an experimental observation of the  Majorana
fermions in two dimensions.
\end{abstract}
\preprint{ALBERTA-THY-7-22}

\maketitle
Majorana  quasiparticle excitations in various condensed matter
systems are in a spotlight of theoretical and experimental
studies for over a decade
\cite{Alicea:2012ux,Beenakker:2013,Elliott:2014iha}. A renown
example of the Majorana quasiparticles are the zero-energy
states  bound to the vortices in a topological superconductor
\cite{Volovik:1999eh,Read:2000} or on the interface between a
topological insulator and a normal superconductor
\cite{Fu:2007moz}. The {\it giant} vortices  of large winding
number $n$ are of particular interest since they host multiple
zero modes \cite{Jackiw:1981ee} and can be used to study
highly nontrivial systems of  interacting Majorana states such
as the Sachdev-Ye-Kitaev model \cite{Pikulin:2017mhj}.  The
vortices with $n>1$ have already  been observed in mesoscopic
semiconductors \cite{Cren:2011} and can be engineered in the
specially designed heterostructures.  Unambiguous
identification of the zero modes in the vortex core poses a
great challenge for the modern experimental techniques
\cite{Sun:2016,Wang:2018}, and the measurement of their spatial
distribution is a promising method to distinguish the true
Majorana states \cite{Zhang:2021}. Thus, the search and design
of the systems with the signature spatial properties of  the
zero modes as well as the theoretical evaluation of their shape
and the local density of states  are of primary interest.
Though the existence and stability of the zero modes are
predicted by the index theorem
\cite{Weinberg:1981eu,Tewari:2007} and can be verified through
a qualitative analysis of the field equations, such an approach
is too coarse to catch  subtle dynamical effects which
significantly affect the structure of the solution. On the
other hand, the brute-force  numerical simulations may be
insufficient to identify the universal properties and
characteristic features of the solution, hence a form of
quantitative analytic approach is mandatory. A systematic
analysis in this case is complicated by nonintegrable nonlinear
nature of the vortex dynamics. In this Letter we present such
an analysis for the case of the giant vortices. It is based on
a novel method of the asymptotic expansion in inverse powers of
the vortex winding number \cite{Penin:2020cxj,Penin:2021xgr}
and provides the analytic solution for almost all of the $n$
zero modes required by the index theorem. Our prediction for
the density of states has a number of remarkable properties
which have been overlooked before and can  be used  to get
compelling experimental evidence of the Majorana vortex states.

The equations for the Majorana zero modes  can be inferred from the
Jackiw-Rossi  theory of charged massless two-component Dirac
fermion in $2+1$ dimensions described by the Lagrange density
\cite{Jackiw:1981ee}
\begin{equation}
{L_{JR}}=i\bar{\psi}\slashed{D}\psi+{1\over 2}
\left(\bar{\psi}\psi^c\phi+\bar{\psi}^c\psi\phi^*\right)\,,
\label{eq::LagrangeJR}
\end{equation}
where $\slashed{D}=\gamma^\mu D_\mu$,
$D_\mu=\partial_\mu+iA_\mu$ is the gauge covariant derivative,
the Dirac matrices reduce to the  Pauli matrices
$\gamma^\mu=(\sigma_3,i\sigma_2,-i\sigma_1)$,
$\psi^c=-i\sigma_1\psi^*$ is the charge conjugate spinor, and
$\phi$ is a scalar field of charge $2$ representing the pair
potential. For the  static zero-energy states the  field
equations for the spinor components read
\begin{equation}
D_\pm\psi^\pm+\phi{\psi^*}^\pm =0\,,
\label{eq::Diraceq}
\end{equation}
where the chiral derivatives are $D_\pm=D_1\pm iD_2$. We are
interested in the solution of Eq.~(\ref{eq::Diraceq}) in the
background of the axially symmetric Abrikosov vortex
\cite{Abrikosov:1956sx} of the winding number $n$, which
implies the following field configuration in polar coordinates
$\phi(r,\theta)=f(r)e^{in\theta}$, $A_\theta=-n a(r)/2$,
$A_r=0$, with  $f(0)=a(0)=0$ and $f(\infty)=f_\infty$,
$a(\infty)=1$. Then the negative chirality equation does not
have a normalizable solution and the $n$ zero modes of positive
chirality can be written as follows
\begin{equation}
\begin{split}
&\xi^+_l={1\over \sqrt{2}}\left(e^{il\theta}\psi^+_l
+e^{i(n-1-l)\theta}\psi^+_{n-1-l}\right)\,,\\
&\eta^+_l={i\over \sqrt{2}}\left(e^{il\theta}\psi^+_l
-e^{i(n-1-l)\theta}\psi^+_{n-1-l}\right)\,,
\label{eq::zmdec}
\end{split}
\end{equation}
where $0\le l\le n/2-1$ for even $n$ and  $0\le l\le (n-1)/2$,
$\eta^+_{(n-1)/2}=0$ for odd $n$. The partial wave amplitudes
satisfy the following equations
\begin{equation}
\begin{split}
& \left({d\over dr}-{l\over r}+{na\over 2r}\right)
\psi_l^++f\psi_{n-1-l}^+=0
\,, \\
&\left({d\over dr}-{n-l-1\over r}+{na\over 2r}\right)
\psi_{n-1-l}^++f\psi_{l}^+=0\,.
\label{eq::BdG}
\end{split}
\end{equation}
After identification of $\psi_{l}$ and $\psi_{n-1-l}$ with the
components of the Nambu spinor,  and of $f$ with the pair
potential the above system reproduces the Bogoliubov-de-Gennes
equations for the  Majorana vortex zero modes of the effective
Dirac fermion at zero chemical potential in the condensed
matter systems (see {\it e.g.} \cite{Chamon:2010ks}).

Let us outline the main idea of our approach. The solution of
Eq.~(\ref{eq::BdG}) requires an explicit form of the vortex
fields. In general the functions $a(r)$ and $f(r)$ can be
systematically obtained only through the numerical calculation
within the self-consistent Bogoliubov-de-Gennes formalism
\cite{Shore:1989}. However, the vortex structure drastically
simplifies for $n\gg 1$. In this limit the vortices evolve into
the {\it thin-wall flux tubes} \cite{Bolognesi:2005zr} with the
nonlinear dynamics confined to a narrow boundary layer  outside
the vortex core \cite{Penin:2020cxj}. The boundary layer depth
is given by the maximal of the magnetic penetration length
$\delta$ and the correlation length $\xi$ of the
superconductor, while the core radius grows with $n$ as
$r_n=2^{3/4}\sqrt{n}\,\zeta$, where $\zeta=\sqrt{\delta\xi}$ is
the geometric average of the scales. Inside and outside the
core the dynamics of the gauge and scalar fields linearizes up
to the corrections exponentially suppressed for large $n$.
Inside the boundary layer the asymptotic vortex solution does
not depend on the winding number and gets the corrections in
powers of $1/\sqrt{n}$. The method to obtain the large-$n$
solution as well as the finite-$n$ corrections based on the
effective field theory idea of  scale separation  has been
developed in Refs.~\cite{Penin:2020cxj,Penin:2021xgr}. In the
present work it is applied to  the analysis of
Eq.~(\ref{eq::BdG}) in  the giant vortex background. Throughout
the paper we consistently use the universal aspects of the
leading order result and neglect the model-dependent
corrections.

It is convenient to decouple the  gauge field  by a field
redefinition
\begin{equation}
\psi_l^+(r)=u_l(r)
\exp\left(-{n\over 2}\int_0^r {a(r')\over r'}{\rm d}r'\right)
\label{eq::psidec}
\end{equation}
and to transform the system Eq.~(\ref{eq::BdG}) into the second
order equation
\begin{equation}
\begin{split}
&\left[{d^2\over dr^2}-\left({n-1\over r}
+{f'\over f}\right){d\over dr}\right.\\
&\left.+{l\over r}\left({n-l\over r}+{f'\over f}\right)-f^2\right]u_l
=0\,,
\end{split}
\label{eq::uleq}
\end{equation}
where $f'=df/dr$. At $r>r_n$  the scalar and gauge fields
exponentially approach their vacuum values and the normalizable
solution of Eq.~(\ref{eq::BdG}) reads
\begin{equation}
\psi^+_l(r)\propto K_{\mu}(r/\sqrt{2}\delta)\,,
\label{eq::ultail}
\end{equation}
where  $K_{\mu}(z)$ is the $\mu$th modified Bessel function
with $\mu=\sqrt{n^2/4-l(n-l)}$, and the relation
$f_\infty=1/\sqrt{2}\delta$ is used. Thus, the solution
exponentially decays outside the core  indicating that the zero
modes are localized in the vortex core or on its boundary.

The field dynamics inside the vortex core is determined solely
by the gauge interaction giving the universal solution
\cite{Penin:2021xgr}
\begin{equation}
\begin{split}
&f(r)=f_0\exp\left[{n\over 2}\left(\ln\left({r^2\over r^2_n}\right)
- {r^2\over r_n^2}+1\right)\right],\\
&a(r)={r^2\over r_n^2}\,,
\label{eq::core}
\end{split}
\end{equation}
where $r<r_n$,  $f_0$ is an inessential integration constant,
and $a(r)$ corresponds to a homogeneous magnetic field.
Though the  pair potential in Eq.~(\ref{eq::core}) is
exponentially suppressed,  it is a singular perturbation
since the order of the Bogoliubov-de-Gennes equations for
vanishing $f$ is reduced. Indeed, for $r<r_n$ the logarithmic
derivative term $f'/f={n/ r}\left(1-{r^2/r_n^2}\right)$ in
Eq.~(\ref{eq::uleq}) is not suppressed and must be kept to
get two regular solutions  at $r=0$. These solutions are
\begin{equation}
\begin{split}
&u_l^{(1)}(r)=r^l\,, \qquad
u_l^{(2)}(r)=r^{2n-l}{\rm E}_{\nu}\left(n r^2/2 r_n^2\right)\,,
\label{eq::ulsol}
\end{split}
\end{equation}
where ${\rm E}_{\nu}(z)$ is the $\nu$th exponential integral
with $\nu = 1+l-n$. The behavior of the second solution at
large $n$ is quite peculiar. For $l<n/2$ it reduces to
$u_l^{(2)}(r)\sim r^l$ {\it i.e.} the two solutions are
degenerate up to the exponentially suppressed terms. For  $l>
n/2$, however,  it transforms into $u_l^{(2)}(r)\sim
r^{2n-l}e^{-n r^2/2r_n^2}$ and is the only solution which gives
an unsuppressed contribution to $\psi_l^+(r)$.  The gauge field
factor in Eq.~(\ref{eq::psidec}) inside the core equals to $e^{-n
r^2/4 r_n^2}$ and for the partial waves in the large-$n$ limit
we finally get
\begin{equation}
\psi^+_l(r)\sim N_l\left\{
\begin{array}{ll}
r^l e^{-n r^2/4 r_n^2} \,,  & l<n/2 \,,\\
r^{2n-l}e^{-3n r^2/4 r_n^2}\,, \quad & l>n/2 \,, \\
\end{array}
\right.
\label{eq::zeromJR}
\end{equation}
where $N_l$ is the normalization factor.
Eq.~(\ref{eq::zeromJR}) describes two groups of approximately
Gaussian peaks. For $l<n/2$ the peaks of the width
$\sigma=r_n/\sqrt{n}=2^{3/4}\zeta$ are centered at
$\bar{r}_l=\sqrt{2l/n} \,r_n$, while for $l>n/2$ the peaks have
the width $\sigma'=\sigma/\sqrt{3}$ and  are centered at
$\bar{r}'_l=\sqrt{(2/3)(2-l/n)}\,r_n$. The solutions
with $l\approx n/2$ are localized inside the boundary layer
where the nonlinear effects are essential and an explicit
analytical solution is not available. At the same time, for
known  functions $f(r)$ and $a(r)$  the solution is given by
\begin{equation}
\psi^+_{n/2}(r)\propto r^{n\over 2}
\exp\left[-\int^r\left({n a(r')\over 2r'}+f(r')\right){\rm d}r'\right]\,,
\label{eq::zeromJRbl}
\end{equation}
where  $n$ is assumed to be even. It describes  a non-Gaussian
peak of an ${\cal O}(\delta)$ width.  Note that the functions
$f(r)$ and $a(r)$ do not depend on $n$ inside the boundary
layer and are known explicitly for Ginzburg-Landau theory in
the integrable limits of  large, critical, or small values of
the Ginzburg-Landau parameter $\kappa=\delta/\xi$
\cite{Penin:2021xgr}.

Let us now discuss the physical nature of the solution
Eq.~(\ref{eq::zeromJR}). For $l<n/2$ it corresponds to the
lowest Landau level states formed by an approximately
homogeneous magnetic field inside the vortex core. Each of
these states encircles  an even number of the  flux quanta.
Hence, only about $n/2$   of the Landau states fit into the
vortex core and are not affected by the pair potential. For
larger $l$ the effect of Andreev reflection  on the
localization of the states increases and for $l\approx n$ it
exceeds the effect of the magnetic field. This follows
{\it e.g.} from a comparison of the  exponential factor in
Eq.~(\ref{eq::psidec}) due to the magnetic field to the
one of  $u_l^{(2)}$ due to the pair
potential. As it has been pointed out for  $l>n/2$ the Andreev
reflection squeezes the Gaussian peaks  by the factor
$\sqrt{3}$ and displaces  them towards the center of the vortex
with the innermost position $\sqrt{2/3}\, r_n$ of the maximal
angular momentum partial wave. Remarkably such a significant
effect is achieved in the region where the pair potential
is exponentially small {\it i.e.} the Andreev reflection in this
case is a {\it long-range} phenomenon. It can be attributed to
the  singular character of the vanishing pair potential limit
for the Bogoliubov-de-Gennes equations discussed above.

\begin{figure}[t]
\begin{center}
\raisebox{35mm}{ $\displaystyle{\rho\over \rho_0}$}
\includegraphics[width=7cm]{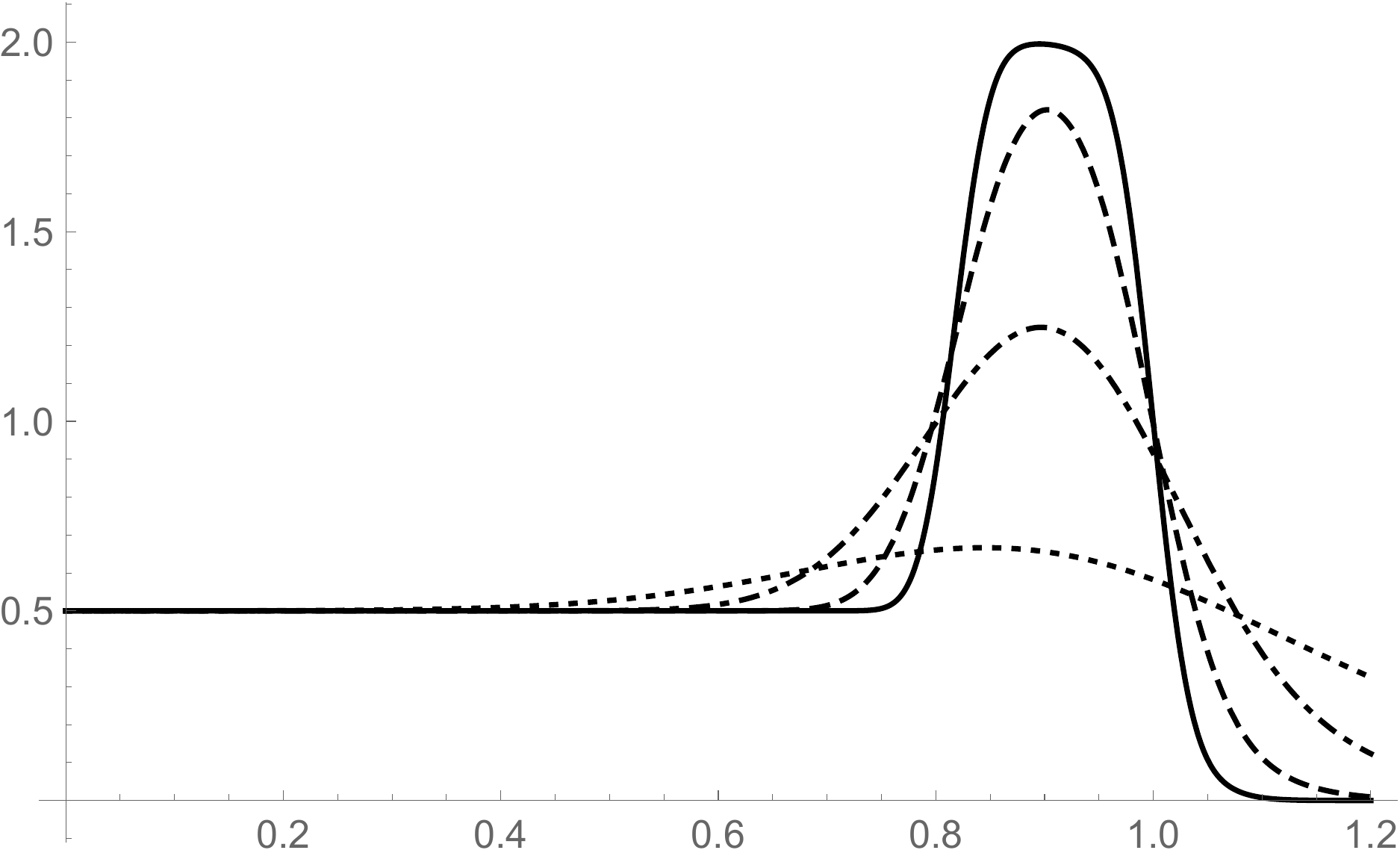}\\
\hspace{40mm}{ $r/r_n$}
\end{center}
\caption{\label{fig::1} The normalized radial density of states
for the Majorana zero modes of a giant vortex with the winding number
$n=4$ (dotted line), $n=16$ (dashed-dotted line), $n=64$ (dashed line),
and $n=256$ (solid line).}
\end{figure}

We are now able to compute the experimentally observable radial
density of states
$\rho(r)=2\pi\sum_{l=0}^{n-1}\left(\psi^+_l(r)\right)^2$ for
the Majorana zero modes.  At  large $n$ the effect of the
poorly approximated $l\approx n/2$ states is negligible and
the  sum converges to the  function
\begin{equation}
\rho(r)\sim \rho_0\left\{
\begin{array}{ll}
{1/ 2} ,  & r/r_n<\sqrt{2/3}\,, \\
& \\[-2mm]
2\,, \quad & \sqrt{2/ 3}<r/r_n<1\,,  \\
\end{array}
\right.
\label{eq::rho}
\end{equation}
where $\rho_0={2n/r_n^2}={1/\sqrt{2}\,\zeta^2}$ does not depend
on $n$. The function  $\rho(r)$ for a few  finite values of the
winding number  is plotted in Fig.~\ref{fig::1}. There the
$l=n/2$ states are approximated by the Gaussian peaks of the
width $\sigma$ centered at $r_n$, which does not significantly
affect the distribution even for the moderate values of $n$. As
we can observe the convergence to the asymptotic result is very
fast for $r<\sqrt{2/3}\, r_n$  and slow for $\sqrt{2/3}\,
r_n<r<r_n$  but  the characteristic shape of the spatial
distribution becomes evident already at $n=4$. Thus, it is more
important to estimate the accuracy of our prediction for the
local density of states at a given moderately large value of
$n$, {\it i.e.} the accuracy of each line in Fig.~\ref{fig::1}.
Inside the core  where most of the states are localized the
accuracy of the method is exponential and for $n=4$ the
estimated error is about  a few percent.  At the core boundary
the accuracy deteriorates  due to the dependence of the
$l\approx n/2$ states on the exact form of the pair potential.
A conservative estimate of the uncertainty  for an individual
state can be done by evaluating the factor $e^{-\int\! f(r')\,{\rm
d}r'}$ in Eq.~(\ref{eq::zeromJRbl}) where the integral runs
over the boundary layer of the depth $\delta$. Approximating
$f(r)$ with $f_\infty/2$ we get the correction factor
$e^{-1/2\sqrt{2}}$ corresponding to a 30\% error. This, however,
affects only the tail of the distribution at $r>r_n$ where the
states with $l\approx n/2$ give the dominant contribution.
Thus, our analysis  is reasonably accurate already for  $n=4$.
The  vortices with such winding number  have already been
observed experimentally \cite{Cren:2011}.

So far we have considered the case  $\kappa={\cal O}(1)$. For
$\kappa\gg 1$ there appears another  condition on the allowed
values of $n$. The method \cite{Penin:2020cxj,Penin:2021xgr}
relies on the scale hierarchy $\delta/r_n\ll 1$. This scale
ratio is proportional to $\sqrt{\kappa/n}$. For $\kappa>n$ the
magnetic field is expelled from the vortex core and the vortex
cannot be considered as a thin-wall flux tube.  At the same
time, the superconductors with the large value of the
Ginzburg-Landau parameter may not be ideal  for the
experimental realization of the giant vortices.  Indeed, the
free energy in this case grows with $n$ as $n^2\ln\kappa$ while
for $\kappa={\cal O}(1)$ it scales as $n/\kappa$
\cite{Penin:2021xgr}. This makes  the giant vortices for large
$\kappa$  much less stable against the decay into the
elementary vortices and, hence,  more difficult to create in an
experiment.

In any case, an experimental realization of the giant Abrikosov
vortices with $n={\cal O}(10)$ may not be an easy task. At the
same time the {\it hard wall} giant vortices of arguably very
large $n$ can be created by a magnetic flux flowing through a
hole in the superconducting film on the surface of the
topological insulator. Such a design has been originally
suggested  in Ref.~\cite{Pikulin:2017mhj} for a physical
realization of the Sachdev-Ye-Kitaev model on the hole
boundary, but it can also be an ideal place for the  study of
two-dimensional Majorana zero modes in the hole interior. The
result Eq.~(\ref{eq::rho}) in this case should be adjusted. The
term $f'/f$ in Eq.~(\ref{eq::uleq}) now gets a very large
positive constant contribution proportional to the ratio of the
Cooper pair chemical potential to the superconductor energy
gap. This effectively  makes the Andreev reflection short-range
so that all the states with $l\gsim n/2$ get localized on the hole
edge. The radial density of states now takes the form
\begin{equation}
\rho(r)\sim {\rho_0\over 2}\left(1+{R\over 2}\delta(r-R)\right)\,,
\label{eq::rhoh}
\end{equation}
where $\rho_0={2n/ R^2}$ and $R$ is the hole radius. The
delta-function in the above equation is in fact an
approximation of the non-Gaussian peak with the width of order
$\delta\ll R$, and $R$ should be taken larger than $r_n$ for a
given $n$ and $\zeta$ to have a stable vortex configuration.

Though the spatial distributions in  Eq.~(\ref{eq::rho}) and
Eq.~(\ref{eq::rhoh}) are quite similar, the physical properties
of the two systems are qualitatively different. For Abrikosov
vortices the parameter $\rho_0$  which defines the average
density of states is $n$-independent and  completely determined
by the intrinsic properties of the superconductor through the
geometric average $\zeta$ of the magnetic penetration and the
correlation length. By contrast, for the hard-wall vortices the
parameter $\rho_0$  is {\it quantized} in the units of
${2/R^2}$ and is proportional to the number of the magnetic
flux quanta. Thus, it can be discreetly changed by the variation
of the applied magnetic field  $B$. The corresponding average
rate of the density variation inside the core evaluates to
\begin{equation}
{d \rho\over d B}=\pi K_J\,,
\label{eq::rhovar}
\end{equation}
where $K_J$ is the Josephson constant (the inverse of the
magnetic flux quantum).

Our  solution Eq.~(\ref{eq::zeromJR}) is qualitatively
different from the existing analysis where the role of the
magnetic field on the formation of the vortex states has been
neglected. This  is indeed justified for an elementary vortex
and for the large values of the Ginzburg-Landau parameter
$\kappa\gg 1$ when the vortex states are predominantly formed
through the Andreev reflection and are localized  near the core
boundary \cite{Caroli:1964,Bardeen:1969,Virtanen:2008}. However,
with the increasing winding number the magnetic flux through
the vortex core grows  and for the giant vortices with $n\gg
\kappa$ the zero modes are formed through a fine interplay
between the magnetic effects  and the long-range Andreev
reflection resulting in a set of the {\it warped} lowest Landau
level states. Note that the effect of the magnetic field on the
delocalization of the zero modes for the elementary vortex
lattice has been recently discussed in
Ref.~\cite{Pacholski:2021}.

To conclude, we have applied an advanced asymptotic method
based on the scale separation and the expansion in inverse
powers of the winding number to find the analytical solution
for the Majorana zero modes of the giant vortices. In the case
of the Abrikosov vortices the solution  reveals a nontrivial
dynamical origin and a simple universal structure,  provided
the vortex winding number exceeds the value of the
Ginzburg-Landau parameter of the superconductor. It is not
sensitive to the form of the pair potential and is completely
determined by a single parameter, the  vortex radius, which can
be directly measured in the experiment. The resulting local
density of states  is confined to the vortex core, where the
non-zero modes are magnetically gapped. The density has a
characteristic profile which can be used as a signature for the
identification of the Majorana zero modes by  scanning electron
microscopy \cite{Zhang:2021}. For the hard-wall giant vortices
in the specially designed heterostructures
\cite{Pikulin:2017mhj} we have found that a half of the zero
modes are pinned to the vortex edge with the other half filling
the vortex core. The dependence of the energy  on the applied
magnetic field, characteristic of  the non-zero modes, can
therefore be used for a clear identification of the Majorana
core states. Moreover, the density of the zero modes is
quantized and changes discreetly under the variation of the
magnetic field with the universal average rate  given by
the Josephson constant. These features establish  the
giant vortices as  an  ideal laboratory for the search of
compelling experimental evidence of the Majorana fermions in
two dimensions.

\vspace{2mm}
\noindent
{\bf Acknowledgments.} We would like to thank Joseph Maciejko
for many useful discussions.  The work of  L.G. was supported
through by NSERC. The work of  A.P. was supported in part by
NSERC and the Perimeter Institute for Theoretical Physics.


\end{document}